\documentclass{ifacconf}
\usepackage{graphicx}      
\usepackage{natbib}        
\begin{document}
\begin{frontmatter}

\title{Driving Style Recognition at First Impression for Online Trajectory Prediction \thanksref{footnoteinfo}} 

\thanks[footnoteinfo]{This work has been submitted to IFAC for possible publication.}

\author[First]{Tu Xu} 
\author[First]{Kan Wu}
\author[First]{Yongdong Zhu}
\author[First]{Wei Ji}

\address[First]{Research Center for Intelligent Transportation, Zhejiang Lab, Hangzhou, China(e-mail: xutu@zhejianglab.com, kanwu@zhejianglab.com, zhuyd@zhejianglab.com, jiw@zhejianglab.com)}

\begin{abstract}                
This paper proposes a new driving style recognition approach that allows autonomous vehicles (AVs) to perform trajectory predictions for surrounding vehicles with minimal data.  Toward that end, we use a hybrid of offline and online methods in the proposed approach. We first learn typical driving styles with PCA and K-means algorithms in the offline part.  After that, local Maximum-Likelihood techniques are used to perform online driving style recognition.  We benchmarked our method on a real driving dataset against other methods in terms of the RMSE value of the predicted trajectory and the observed trajectory over a 5s duration.  The proposed approach can reduce trajectory prediction error by up to 37.7\% compared to using the parameters from other literature and up to 24.4\% compared to not performing driving style recognition.

\end{abstract}


\end{frontmatter}
\section{Introduction}
Predicting the trajectory of surrounding traffic participants is a critical predecessor task for autonomous driving motion planning. For safety purposes, trajectory prediction needs to be precise and computationally efficiently \citep{RN1}, making this task difficult. Vehicles’ driving behavior is highly uncertain, and it could be impacted by various observable and unobservable factors, including traffic signs, road environment, and drivers’ psychological and physiological factors, which can be referred to as ``driving style''. An experienced driver would form first impressions of surrounding vehicles’ driving style, usually after a short period of observation, even at a glance, and the driving style recognition helps the drivers to categorize their prediction on neighbor vehicles’ trajectory to help do their motion planning.

Driving style is a complex concept \citep{RN2, RN3}. It is a common approach to study driving style from speed profile by interpreting driving activities such as acceleration, deceleration, turning and lane changing \citep{RN4}. Common driving style recognition methods include rule-based algorithms to allocate driving behavior into groups based on predefined thresholds \citep{RN5}, fuzzy-logic regression \citep{RN6}, unsupervised machine learning algorithms such as principal component analysis (PCA) \citep{RN7} and K-means clustering \citep{RN8}, supervised machine learning algorithms such as k-nearest neighbors (kNN) \citep{RN9}, support vector machine (SVM) \citep{RN10} and artificial neural networks (ANN) \citep{RN11}.

In terms of trajectory prediction, there are generally two types of approaches, including data-driven and rule-based methods. Data-driven methods such as Gaussian mixture models and neural networks \citep{RN12, RN13, RN14} often have the expressive capability to capture the subtle differences among various driving styles. However, the data-driven approaches are like black-box and lack interpretability, and they could sometimes make unrealistic predictions and threaten ego vehicle’s driving safety \citep{RN15}. Moreover, data-driven methods are usually more computing-intensive, requiring more input data and longer computing time than rule-based methods. For autonomous driving under single vehicle intelligence (in contrast to connected vehicles, which can get beyond-visual-range information through vehicle-to-vehicle or vehicle-to-road communications), on-board sensors can hardly observe the exact neighboring vehicle for a long period, except the leading vehicle in adaptive cruise control mode. Thus, AVs must make online trajectory predictions with a minimal amount of observation data.

Rule-based methods apply driver models, e.g., the IDM \citep{RN16}, the Gipps’ Model \citep{RN17}, and the Newell’s Model \citep{RN18}, to fit observed car-following and lane-changing behaviors and to make trajectory prediction. Since rule-based models take advantage of expert knowledge, they can make relatively robust trajectory predictions even with underrepresented data \citep{RN19}. The precision of rule-based models relies on the parameter estimation of driver models. Model parameters can be chosen either offline or online. Offline approaches can use a relatively larger amount of data to generate a set of average parameters to portray typical driving style, and there is no computing time constraint. Online approaches use real-time observation data to calibrate model parameters better to fit individual characteristics \citep{RN20}. According to a survey paper \citep{RN21}, only a small fraction of research attempted to calibrate car-following models.

This paper proposes a new approach as a hybrid of offline and online methods, allowing us to perform trajectory prediction with limited observation data. Our approach can take advantage of an arbitrary amount of historical data to learn typical driving styles so that it can form a first impression of a newly observed vehicle’s driving style with a significantly shorter observation period, just like an experienced driver. Moreover, it will dynamically correct its driving style recognition and accumulate sensor data about the exact surrounding vehicle to perform a precise trajectory prediction at relatively low computing costs.

\section{Methodology}
\begin{figure}
\begin{center}
\includegraphics[width=8.4cm]{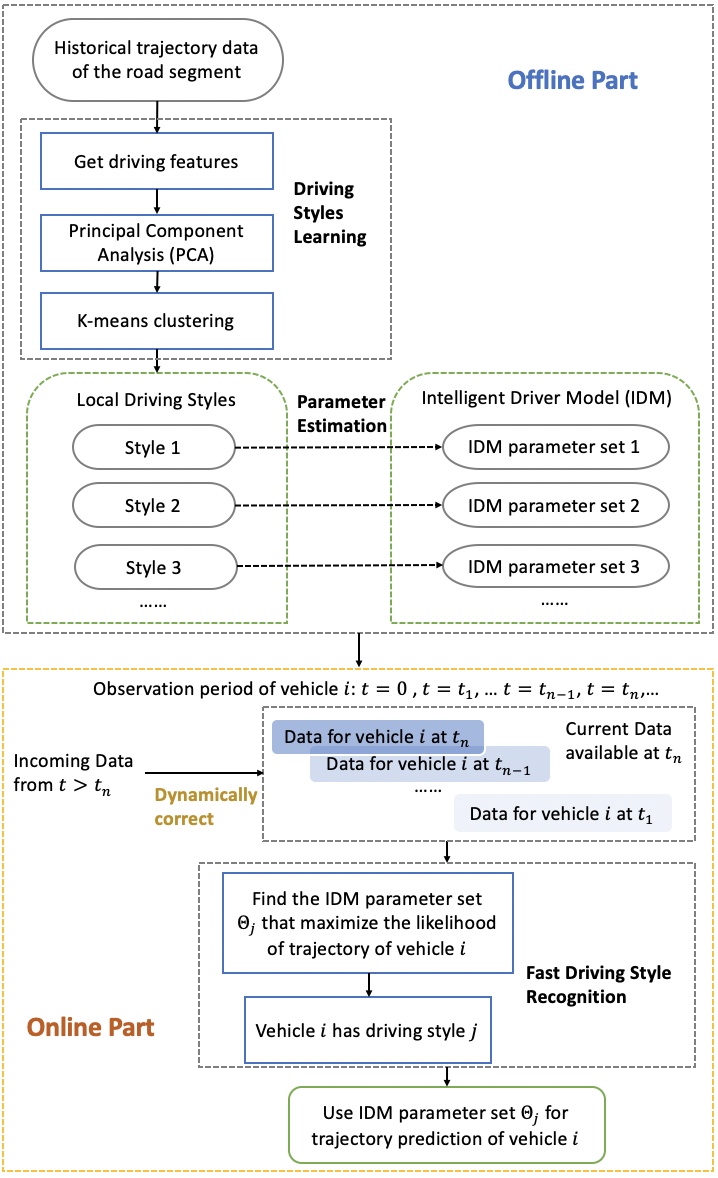}    
\caption{Framework of the proposed approach for driving style recognition and trajectory prediction.
} 
\label{framework}
\end{center}
\end{figure}

The proposed trajectory prediction approach consists of offline and online parts.  For the offline part, an arbitrary amount of historical trajectory data are used to learn typical driving styles.  Then, the Intelligent Driving Model (IDM) car-following parameters for each driving style are estimated.  The clustering analysis results and the IDM parameter value sets help AVs to form a first impression of a newly observed vehicle's driving style with a significantly shorter observation period in the online part.  For the online part, given the trajectory data of a surrounding vehicle $i$, one can find the IDM parameter value set $\Theta_j$ that best fits the trajectory.  The driver is then classified into the corresponding driving style cluster $j$.  One thing to note is that, along with the accumulation of trajectory data of vehicle $i$, the approach can dynamically correct its driving style recognition.  The framework of the proposed approach is shown in Fig. ~\ref{framework}.

\section{Experiments}
\subsection{Data}
We evaluate the performance of our driving style recognition methods on demonstration data from a real-world dataset, namely the reconstructed I80-1 dataset (from 4.00 p.m. to 4.15 p.m.) \citep{MONTANINO201582, PUNZO20111243}, which provides the X, and Y coordinates of each vehicle at 10Hz. 
 In this paper, we perform driving style recognition only based on car-following features of vehicles.  Toward that end, car-following pairs (leader-follower) were extracted from the original dataset.  In each car-following pair, the car-following driving condition between the leader and the follower lasts for at least 15 seconds without any lane changing or interference from other vehicles.  We extracted 833 car-following pairs from the inner four lanes in total.  We use 80\% of them (666 pairs) in the offline part (referred to as dataset 1 hereafter) and 20\% of them (167 pairs) in the online part (referred to as dataset 2 hereafter).

\subsection{The offline part: typical driving styles learning}
We need ego vehicle profiles such as speed and acceleration to identify the driving style of a single vehicle.  On the other hand, we need to consider the interaction between the ego vehicle and its leader.  We select 13 features as characteristic indexes for vehicle driving style learning, as shown in Table \ref{feature}.
\begin{table}[h!]
\begin{center}
\caption{Driving style characteristic indicators}\label{feature}
\begin{tabular}{cl}
No. &  Characteristic indicators  \\\hline
$X_1$ & Maximum speed (m/s) \\
$X_2$ & Mean speed (m/s) \\
$X_3$ & Standard deviation of speed (m/s) \\
$X_4$ & Maximum acceleration (m/s$^2$) \\
$X_5$ & Minimum acceleration (m/s$^2$) \\
$X_6$ & Mean acceleration (m/s$^2$) \\
$X_7$ & Standard deviation of acceleration (m/s$^2$) \\
$X_8$ & Maximum gap(m) \\
$X_9$ & Minimum gap (m) \\
$X_{10}$ & Average gap (m) \\
$X_{11}$ & Standard deviation of the gap (m) \\
$X_{12}$ & Average speed difference (m/s)\\
$X_{13}$ & Standard deviation of the speed difference (m/s) \\
\end{tabular}
\end{center}
\end{table}

We extracted the above 13 features from 15 seconds of car-following data in each car-following pair from dataset 1.  After the feature selection, dimension reduction is carried out on the 13 selected feature indexes such that the low-dimensional representation retains some meaningful properties of the original data.  This paper uses Principal Component Analysis (PCA) \citep{abdi2010principal} as the dimension reduction method. PCA performs a linear mapping of the data to a lower-dimensional space in such a way that the variance of the data in the low-dimensional representation is maximized.  PCA calculation was performed on the trajectory data, and the explained variance ratio by the first five principal components can be found in Table \ref{pca}.  As seen in the table, selecting the first two principal components could represent most of the information (greater than 90\%) about vehicle car-following behaviors.

\begin{table}[h!]
\begin{center}
\caption{Explained variance ratio by the first five principal components}\label{pca}
\begin{tabular}{ccc}
Principal Component &  Explained Variance Ratio & Accumulated  \\\hline
PC1 & 0.862 & 0.862 \\
PC2 & 0.083 & 0.945 \\
PC3 & 0.042 & 0.987  \\
PC4 & 0.004 & 0.991 \\
PC5 & 0.003 & 0.994  \\
\end{tabular}
\end{center}
\end{table}

\begin{figure}
\begin{center}
\includegraphics[width=7.0cm]{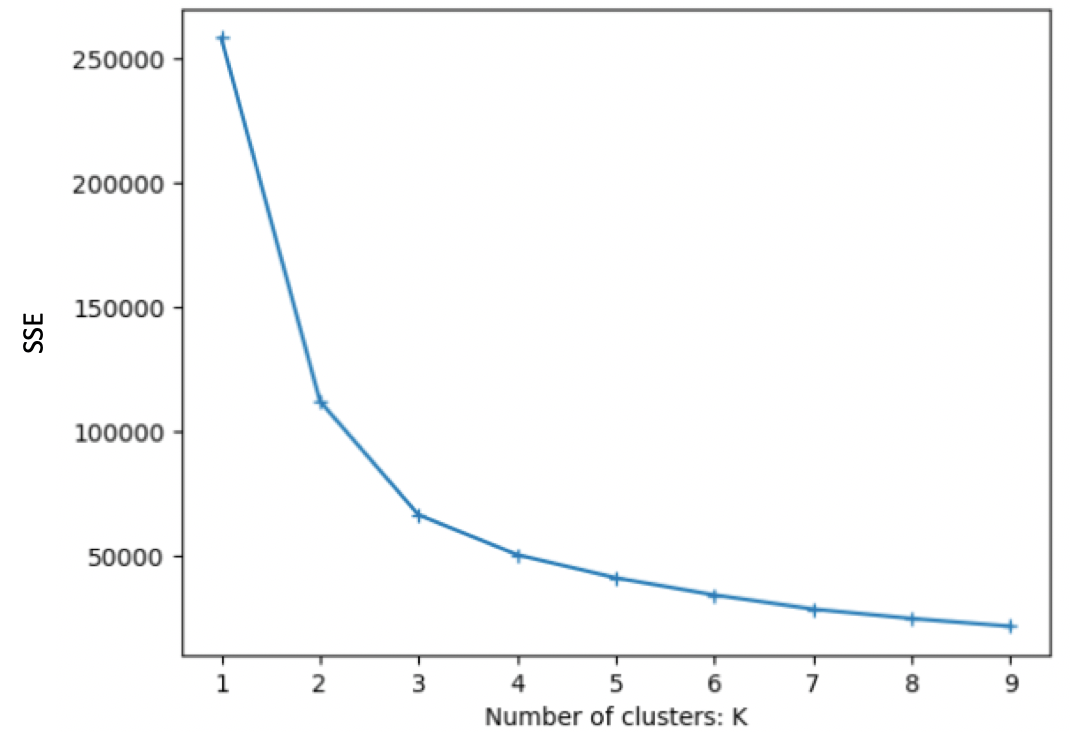}    
\caption{The relationship between SSE value and the number of clusters
} 
\label{elbow}
\end{center}
\end{figure}

\begin{figure}
\begin{center}
\includegraphics[width=8.4cm]{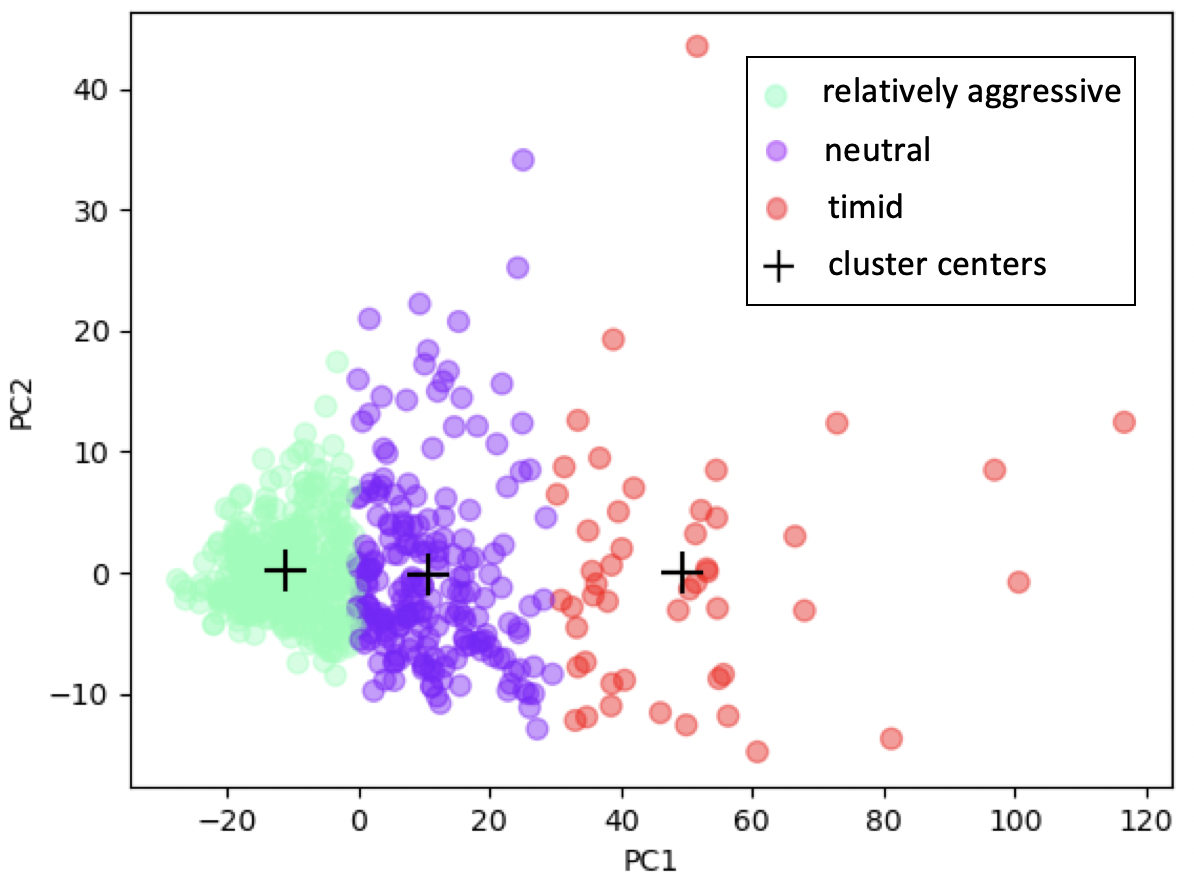}    
\caption{Driving styles clustering results. The method for driving style labeling (``relatively aggressive", ``neutral," and ``timid") is explained in the next subsection.
} 
\label{kmeans}
\end{center}
\end{figure}

After the dimension reduction process, the K-means algorithm was used to classify typical vehicle driving styles\citep{k-means}.  Firstly,  To determine the optimal number of clusters, we have to select the value of $K$ at the “elbow,” which is the point after which the Error Sum of Squares (SSE) within the cluster starts decreasing linearly.  According to Fig. \ref{elbow}, $K=3$ was selected as the optimal cluster number in this paper.  The obtained clustering results are shown in Fig. \ref{kmeans} and Table \ref{tb:kmeans}. Physical meanings of PC1 and PC2 are obscure, but we will explain how we label driving styles (``relatively aggressive," ``neutral," and ``timid") for each driving style cluster in the following subsection.

\begin{table}[h!]
\begin{center}
\caption{Driving styles K-means clustering results}\label{tb:kmeans}
\begin{tabular}{cccc}
Cluster Index &  Driving Style & Sample Size & Proportion\\\hline
0 & neutral & 209 & 31.4\% \\
1 & relatively aggressive & 409 & 61.4\% \\
2 & timid & 48 & 7.2\% \\
\end{tabular}
\end{center}
\end{table}

\subsection{The offline part: car-following parameter estimation for typical driving styles}
To label clusters with driving styles, many methods in the literature \citep{hekmeans9922336, RN10} calculate the mean of features for each cluster.  However, in this way, these works cannot make trajectory predictions based on the clustering results.  

To fill this gap, this paper estimate the car-following parameter values for each driving style based on the Intelligent Driver Model (IDM) \citep{RN16}.  The form of the IDM acceleration is:

\begin{equation} \label{eq:IDM}
\begin{array}{cc}
a_{IDM} &= a_{m}[1-(\frac{v}{v^*})^{\delta}-(\frac{d^*}{d})^2],\\
 d^{*}&=d_{min}+vT-\frac{v(v_L-v)}{2\sqrt{a_{m}b_{comf}}},
\end{array}
\end{equation}
where $\delta$ ($\delta=4$ in this paper) is the free-drive exponent, $v$ is the current speed of the follower, $v_L$ is the current speed of the leader, $d$ is the current gap between the follower and its leader, $d^*$ is the desired gap, and other parameters are to be estimated based on the trajectory data, as seen in Table \ref{tbIDM}:

\begin{table}[hb]
\begin{center}
\caption{IDM parameters to be estimated}\label{tbIDM}
\begin{tabular}{ccc}
Parameter Name &  Physical Meaning & Unit\\\hline
$v^*$ & desired velocity & m/s \\
$T$ & desired time headway & s \\
$d_{min}$  & minimum spacing & m \\
$a_{m}$  & maximum vehicle acceleration & m/s$^2$ \\
$b_{comf}$ & comfortable braking deceleration & m/s$^2$ \\
\end{tabular}
\end{center}
\end{table}

\begin{figure}
\begin{center}
\includegraphics[width=8.4cm]{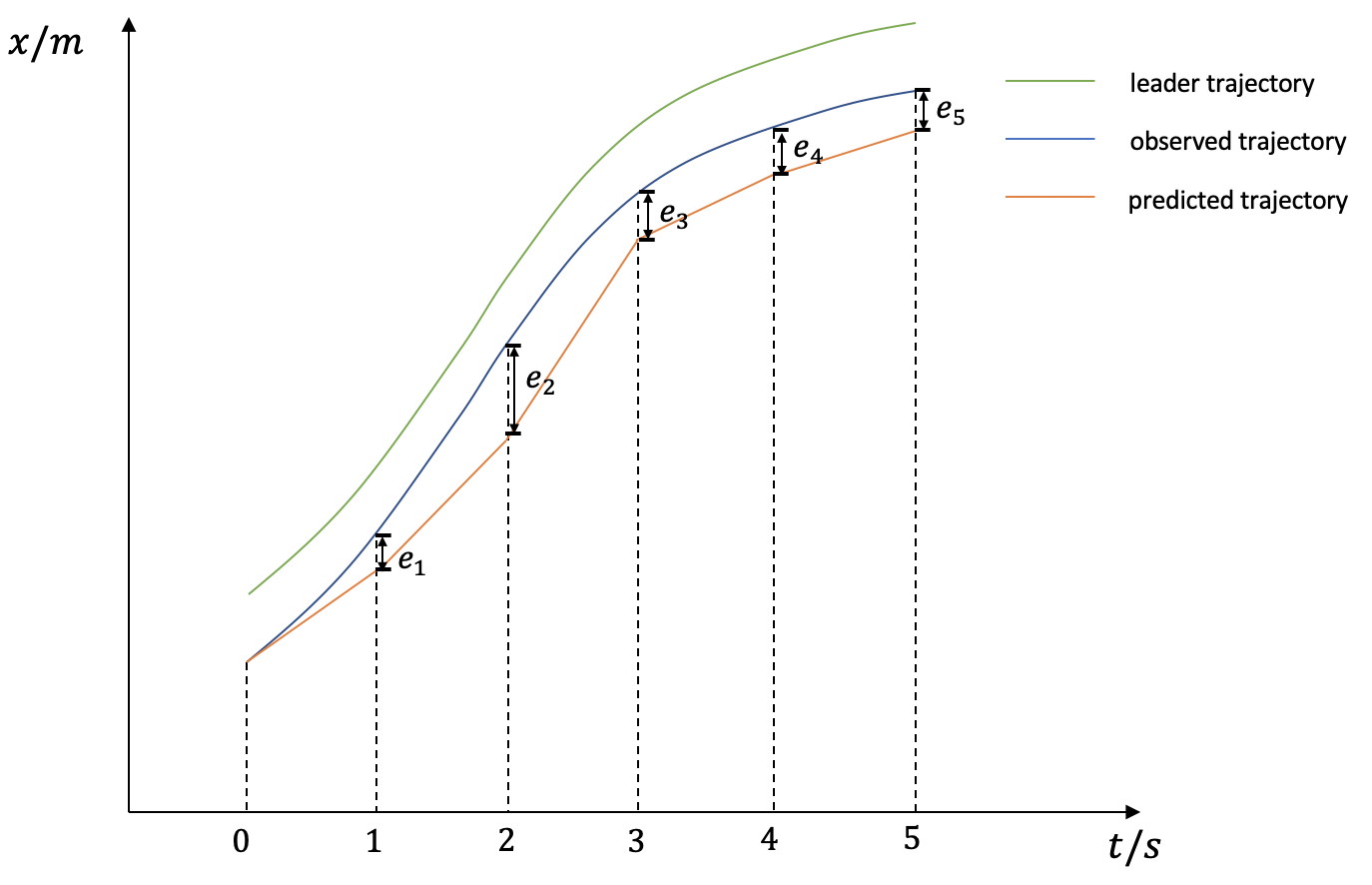}    
\caption{The calculation method of RMSE for a single vehicle over a 5s duration
} 
\label{fig:rmse}
\end{center}
\end{figure}

For each vehicle, we use the simulation of the IDM model to make trajectory predictions over a 5s duration, which then are compared with the observed trajectories by formulating objective functions in terms of Root Mean Square Error (RMSE):

\begin{equation} \label{eq:rmse}
S(\Theta)=\sqrt{\sum_{t=1}^{5}e_t^2(\Theta)/5}=\sqrt{\sum_{t=1}^{5}(\hat{y_t}(\Theta)-y_t)^2/5},
\end{equation}

where $e_t$ is the error between the predicted trajectory $\hat{y}(\Theta)$ and the observed trajectory $y$ at time $t$, and the value is a function of IDM parameters $\Theta = (v^*, T, d_{min}, a_{m}, b_{comf})$.  The calculation method is shown in Fig. \ref{fig:rmse}.

We can calculate the average RMSE $\bar{S}(\Theta)$ for vehicles in the single driving style cluster.  Minimizing $\bar{S}(\Theta)$ reduces the parameter estimation problem to a multivariate nonlinear optimization problem.  The parameter estimation results for three driving style clusters are shown in Table \ref{tbpars}.  The parameter estimation result, $\Theta_{aggre}$, is also calculated if all vehicles are classified into only one cluster.

\begin{table}[hb]
\begin{center}
\caption{IDM parameters for each cluster}\label{tbpars}
\begin{tabular}{cccccc}
Cluster Index & $v^*$ & $T$ &$d_{min}$ & $a_{m}$ & $b_{comf}$ \\\hline
0 & 34.7 & 1.0 & 2.9 & 0.5 & 1.5\\
1 & 35.0 & 1.0 & 0.1 & 0.4 & 1.5 \\
2  & 18.5 & 1.9 & 4.5 & 0.4 & 1.4 \\
aggregate & 19.0 & 1.0 & 0.3 & 0.4 & 1.4 \\
\end{tabular}
\end{center}
\end{table}

From the IDM parameters, we can observe that vehicles in``Cluster 2" show a timid driving style, interpreted by small $v^*$, large $T$, and big $d_{min}$.  Vehicles in ``Cluster 1" and ``Cluster 0" have similar IDM parameters but those in ``Cluster 1" show relatively aggressive driving behaviors, shown by the smaller $d_{min}$ value.  Based on the analysis, we label the driving style of ``Cluster 0" as ``neutral," ``Cluster 1" as ``relatively aggressive," and ``Cluster 2" as ``timid."

\subsection{The online part: fast driving style recognition}
This subsection presents two methods to recognize a vehicle's driving style with limited trajectory data.  The effects of the proposed methods will be compared against benchmark models in terms of the RMSE value of the predicted trajectory and the observed trajectory over a 5s duration.

\subsubsection{method 1}
The idea of method 1 is to repeat the driving style learning process in the offline part.  We calculate the same features listed in Table \ref{feature} from the online vehicle trajectories.  Then the PCA process used in the offline part is applied to reduce the dimension of the feature data.  After that, we calculate the distances between this vehicle's first two principal components (PC1, PC2) and the three cluster centers.  The vehicle belongs to the closest cluster center and has the corresponding driving style.  

\begin{figure}
\begin{center}
\includegraphics[width=7.0cm]{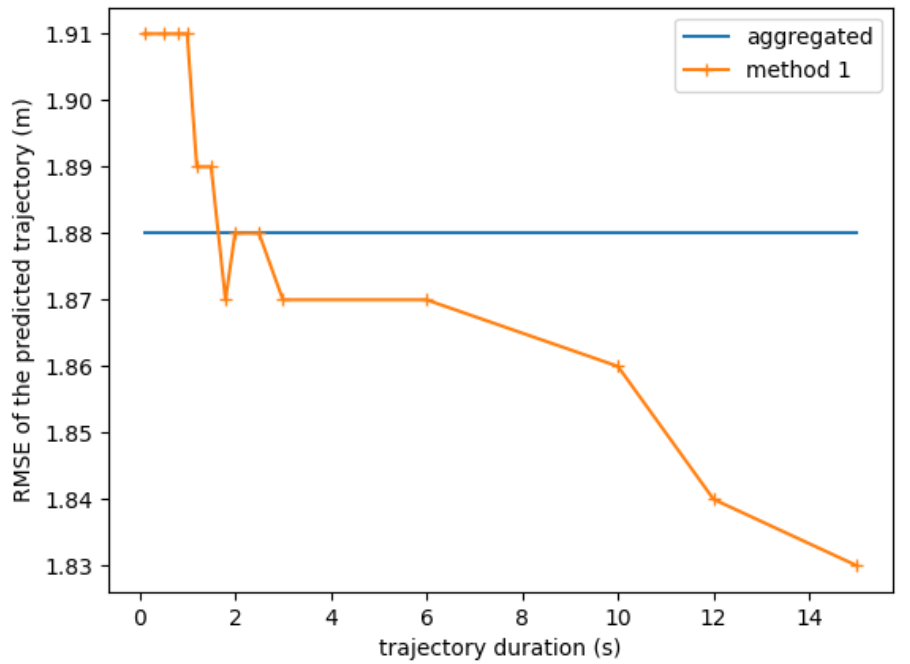}    
\caption{The relationship between the prediction errors (RMSE) and data available ($t_{dur}$) for online driving style recognition method 1
} 
\label{fig:method1}
\end{center}
\end{figure}

We test method 1 with dataset 2.  For each car-following pair in the dataset, trajectory $\textbf{x}$ from $t=0$ to $t=t_{dur}$ is selected.  $\textbf{x}(t_i)$ not only represents the vehicle's location but includes all information needed for extracting the car-following features.  If the driving style of vehicle $i$ is recognized as $k \in \{0, 1, 2\}$, the IDM parameter set $\Theta_{k}\in \{\Theta_{0}, \Theta_{1}, \Theta_{2}\}$ will be used to make predictions for vehicle $i$.  The results show that ``driving style recognition method 1'' can reduce the prediction error by up to 19.7\% (2.28m to 1.83m) compared to simply using parameter values $\Theta_{lit} = (33.3\textnormal{m/s}, 2.0\textnormal{s}, 1.6\textnormal{m}, 0.73\textnormal{m/s}^2, 1.67\textnormal{m/s}^2)$ recommended in the literature \citep{recommendedIDM}, and up to 2.7\% (1.88m to 1.83m) compared to use the estimated aggregated parameters $\Theta_{aggre} = (19.0\textnormal{m/s}, 1.0\textnormal{s}, 0.3\textnormal{m}, 0.4\textnormal{m/s}^2, 1.4\textnormal{m/s}^2)$.  However, method 1 requires a relatively long trajectory to perform well, as shown in Fig. \ref{fig:method1}.  A minimum of 2 seconds of trajectory is required for cluster method 1 to outperform making trajectory predictions with aggregated estimated parameter values $\Theta{aggre}$.

\subsubsection{method 2}
Since AVs are required to make online trajectory predictions with minimal observation data, we propose the second clustering method based on local Maximum-Likelihood calibration \citep{TREIBER2013922}.  In this local method, the acceleration generated by the IDM model $a_{pred}$ is compared against the observed data $a_{real}$, separately for each data point, through maximum-likelihood techniques \citep{Hoogendoorn}.  Toward that end, we assume the noise of the IDM model, $\epsilon$, follows a Normal distribution with a standard deviation of $\sigma$,

\begin{equation} \label{eq:stochasticIDM}
\begin{array}{ll}
a_{pred} = a_{IDM}(\Theta) + \epsilon,\\
\epsilon \sim iid \ N(0, \sigma^2),\\
\end{array}
\end{equation}

such that for each vehicle, we can calculate the log-likelihood function as the sum of each time step $i$, 

\begin{equation} \label{eq:mle}
\begin{array}{ll}
L(\Theta, \textbf{x}) = \sum_{i=1}^{n}ln(\frac{e^{-\frac{(a_{real}(t_i)-a{IDM}(\Theta)(t_i))^2}{2\sigma^2}}}{\sqrt{2\pi}\sigma}),
\end{array}
\end{equation}

where $n$ represents the total data points for driving style recognition from a single vehicle.  The sampling frequency of the I-80 data is 10Hz, so if we have 2 seconds of trajectory data, we will have $n=2 \times 10 = 20$ data points for driving style recognition.  $\textbf{x}=(\textbf{x}(t_1), \textbf{x}(t_2)...)$ represents the trajectory data used in equation \ref{eq:mle}.  $\textbf{x}(t_i)$ not only represents the vehicle's location but includes all exogenous variables needed for the IDM model at time $t_i$.

In traditional vehicle parameter estimation methods, the parameter values for each vehicle are found by maximizing the log-likelihood function $\hat{\Theta} = \textnormal{arg max} L(\Theta, \textbf{x})$.  However, the method in the literature is computing-intensive, which is not ideal for online trajectory prediction.  In this paper, we already have three prototype parameter sets: $\Theta_0 = (34.7\textnormal{m/s}, 1.0\textnormal{s}, 2.9\textnormal{m}, 0.5\textnormal{m/s}^2, 1.5\textnormal{m/s}^2)$ for cluster 0, $\Theta_1= (35.0\textnormal{m/s}, 1.0\textnormal{s}, 0.1\textnormal{m}, 0.4\textnormal{m/s}^2, 1.5\textnormal{m/s}^2)$ for cluster 1, and $\Theta_2= (18.5\textnormal{m/s}, 1.9\textnormal{s}, 4.5\textnormal{m}, 0.4\textnormal{m/s}^2, 1.4\textnormal{m/s}^2)$ for cluster 2, which simplify the parameter estimation problem to finding the vehicle prototype parameter set $k$ that maximizes the log-likelihood:

\begin{equation} \label{eq:mlecluster}
\begin{array}{l}
k = \textnormal{arg max} L(\Theta_k, \textbf{x}).
\end{array}
\end{equation}

\begin{figure}
\begin{center}
\includegraphics[width=8.4cm]{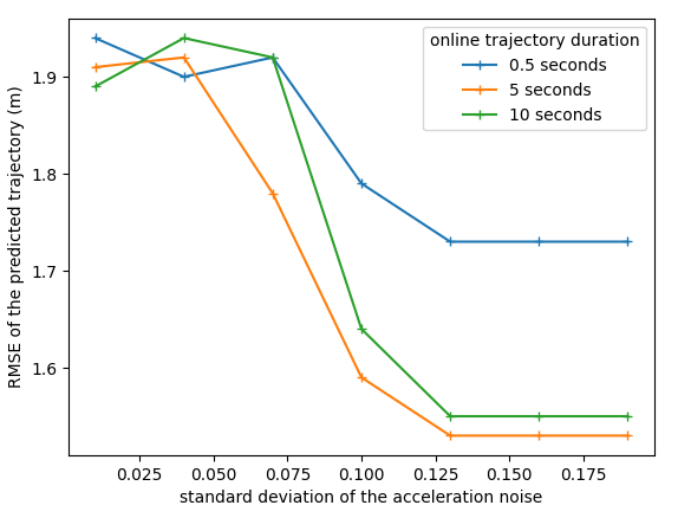}    
\caption{The relationship between the prediction errors (RMSE)  and  $\sigma$ for driving style recognition method 2
} 
\label{fig:sigma}
\end{center}
\end{figure}

Method 2 is tested with dataset 2, with different values of $\sigma \in [0, 0.2\textnormal{m/s}^2]$ and different trajectory duration.  For each car-following pair, we select trajectory $\textbf{x}=(\textbf{x}(0), \textbf{x}(0.1)...\textbf{x}(t_{dur}))$ for driving style recognition.   We run experiments to obtain the relationship between RMSE and $\sigma$, with three different values of $t_{dur}$.  As seen in Fig \ref{fig:sigma}, this driving recognition method achieves the best performance when $0.13\textnormal{m/s}^2 \le \sigma \le 0.2\textnormal{m/s}^2$.  The result is reasonable because human driving behavior is inherently stochastic.

\begin{figure}
\begin{center}
\includegraphics[width=8.4cm]{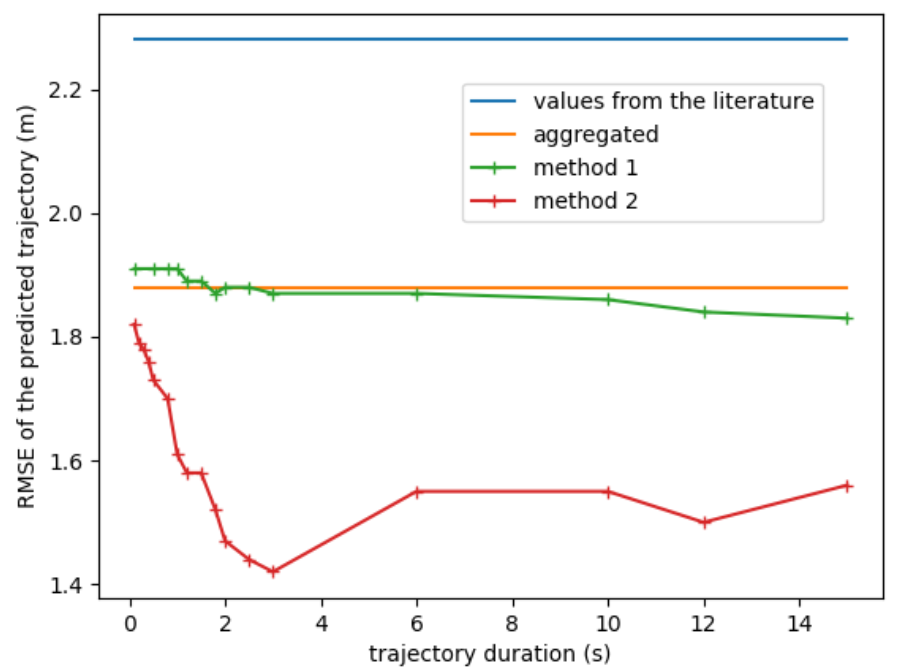}    
\caption{The relationship between the prediction errors (RMSE)  and data available ($t_{dur}$) for online driving style recognition
} 
\label{fig:method2}
\end{center}
\end{figure}

In the following experiment, we choose $\sigma=0.15\textnormal{m/s}^2$.  The driving style recognition method is then run with different duration of trajectories. A comparison between the two methods proposed in this paper and benchmark methods is shown in Fig. \ref{fig:method2}. As seen in the figure,  method 2 reduces the prediction error by up to 37.7\% (2.28m to 1.42m) compared to using parameter values $\Theta_{lit}$, and up to 24.4\% (1.88m to 1.42m) compared to $\Theta_{aggre}$.  Moreover, method 2 outperforms method 1 under both short-trajectory and long-trajectory conditions.  Even with only 0.1 seconds of trajectory data available, the prediction error is smaller than with other methods.

\section{Conslusion}
This paper proposes a new hybrid of offline and online methods, allowing AVs to perform trajectory predictions for surrounding vehicles with limited observation data.  We test this approach on a real driving dataset. 
 We first classify vehicles into three driving styles and estimate the IDM parameter for each style.  After that, we used local Maximum-Likelihood techniques to perform driving style recognition with real-time observation data.  We benchmarked our method against other methods in terms of the RMSE value of the predicted trajectory and the observed trajectory over a 5s duration.  The results show that the proposed approach can reduce the prediction error by up to 37.7\% compared to using the parameters from other literature and up to 24.4\% compared to not performing driving style recognition.

While we perform online driving style recognition, we keep accumulating observation data for a single vehicle in this approach.  However, a vehicle's driving style may change over time.  In future work, we will investigate how to choose a suitable "window" of the trajectory data for driving style recognition.  Moreover, While this study only uses observation data from vehicles, future work will also extend the methods with the help of sensor data from roadside units.  More datasets will be used to assess the performance of the proposed approach in future work.  Finally, the framework of the proposed approach shown in Fig. \ref{framework} is very flexible  so that one can incorporate newer car-following models and other machine-learning algorithms to make online trajectory predictions.

\bibliography{ifacconf}

\end{document}